\documentclass[twocolumn,prl,aps,showpacs]{revtex4}
\usepackage{graphicx}

\begin{document}

\title{Unified demonstration of nonlocality at detection and the Michelson-Morley result \\by a single-photon experiment.}

\author{Antoine Suarez}
\address{Center for Quantum Philosophy, P.O. Box 304, CH-8044 Zurich, Switzerland\\
suarez@leman.ch, www.quantumphil.org}

\date{October 2, 2010}

\begin{abstract}

An experiment demonstrating nonlocality at detection is proposed. A single-photon Michelson-Morley experiment can be performed with the same setup, under exactly the same conditions as well. Hence, the two experiments are ``loophole free'' to the same extent. It is argued that both quantum theory and relativity share the same experimental basis and derive from the same principles.

\end{abstract}


\maketitle

\begin{figure}[t]
\includegraphics[width=80 mm]{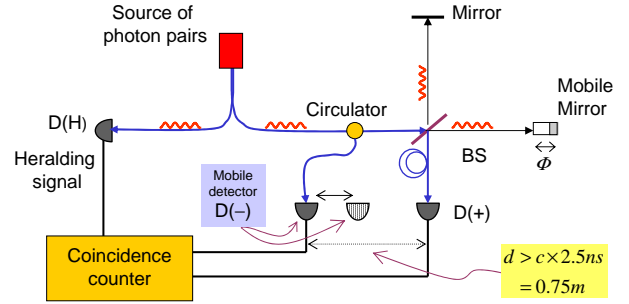}
\caption{Experiment demonstrating nonlocality at detection; the same setup makes it possible to realize a single-photon Michelson-Morley experiment. See text.}
\label{f2}
\end{figure}

Consider the single-particle interference experiment sketched in Figure \ref{f2} using a Michelson-Morley interferometer: A source produces pairs of photons and one of them is used for heralding, i.e. to signaling the presence of a photon in the interferometer and opening the counting gate, as indicated in \cite{herald}. The other photon enters the interferometer through the beam-splitter BS and, after reflection in the mirrors, leaves through BS again and gets detected. The length of one of the arms can be changed at will by means of a mobil-mirror.

Such interference experiments can be considered the entry into the quantum world. With sufficiently weak intensity of light, only one of the two detectors clicks: either D($+$) or D($-$) (\emph{photoelectric effect}). Nevertheless, for calculating the counting rates of each detector one must take into account information about the two paths leading from the laser source to the detector (\emph{interference effect}). If $a \;\in\{+1,-1\}$ labels the detection value, according to whether D($+$) or D($-$) clicks, the probability of getting $a$ is given by:

\begin{footnotesize}
\begin{eqnarray}
P(a)=\frac{1}{2}(1+ a \cos \mathit\Phi)
\label{Pa}
\end{eqnarray}
\end{footnotesize}

\noindent where $\mathit\Phi=\omega\tau$ is the phase parameter and $\tau=\frac{l-s}{c}$ the optical path.

On the other hand the same experiment (Figure \ref{f2}) can be considered the entry into relativity as well, because the Michelson-Morley experiment (the basis of relativity) is an interference experiment.

According to standard quantum mechanics which detector clicks (the outcome) becomes determined at the detection. By contrast, the ``pilot or empty wave'' (de Broglie-Bohm) picture \cite{db}, assumes  that the outcomes become determined at the beam-splitter BS1.

The standard view has the following noteworthy implication: The outcome is decided by a choice on the part of nature when the information about the two paths reaches the detectors. To this aim, there is sort of coordination between D($+$) and D($-$), no matter how far away from each other these detectors are. This implies in principle that interference phenomena cannot be explained by local causality, that is signals propagating between the detectors with velocity $v\leq c$, supposed energy propagation is upper bounded by the velocity of light $c$.

This nonlocality assumption was already implicit in the idea of the \emph{wave function collapse} (also called Copenhagen interpretation), and provoked Einstein's in the Solvay Congress (1927), leading thereafter to the EPR controversy (1935). So, historically, nonlocality at detection appears before Bell's nonlocality and begets it to some extent. In fact, most physicists share the standard view and subconsciously assume nonlocality at detection. One may therefore wonder why this feature has not been specifically addressed by experiment so far.

\textbf{Experiment}. The principle of nonlocality at detection can be tested by the experiment proposed in Figure \ref{f2}. We introduce the following notations:

$P(1,0)$: probability of getting a count in detector D($+$) and no count in D($-$)the other;
$P(0,1)$: probability of getting no count in D($+$)the other and one count in detector D($-$);
$P(1,1)$: the probability of getting one count in either detector;
$P(0,0)$: probability of getting no count in any detector.

Suppose the detectors are activated in such a way that they could signal to each other during a time window of $\Delta t=2.5 ns$ (like in the setup of Reference \cite{herald}). Suppose further that one moves D($-$) to a distance $d$ such that $d>c\Delta t=0.75 m$ (signaling threshold). Then, if one assumes that coordinated behavior of the detectors always requires signaling at $v\leq c$ (local theory), each detector will click independently of the other.

Consequently, according to the local theory, if $d\leq c \Delta t=0.75 m$ the probabilities are given by:

\begin{footnotesize}
\begin{eqnarray}
&&P(1,1)=P(0,0)=0\nonumber\\
&&P(1,0)=\frac{1}{2}(1+\cos \mathit\Phi)\nonumber\\
&&P(0,1)=\frac{1}{2} (1-\cos \mathit\Phi)
\label{P}
\end{eqnarray}
\end{footnotesize}

\noindent and if if $d>c\Delta t=0.75 m$ the probabilities are given by:

\begin{footnotesize}
\begin{eqnarray}
&&P'(1,1)=P'(0,0)=\frac{1}{4}(1- \cos^2 \mathit\Phi) \nonumber\\
&&P'(1,0)=\frac{1}{4}(1+ 2\cos \mathit\Phi+\cos^2\mathit\Phi) \nonumber\\
&&P'(0,1)=\frac{1}{4}(1- 2\cos \mathit\Phi+\cos^2\mathit\Phi)
\label{P'}
\end{eqnarray}
\end{footnotesize}

\begin{figure}[t]
\includegraphics[width=80 mm]{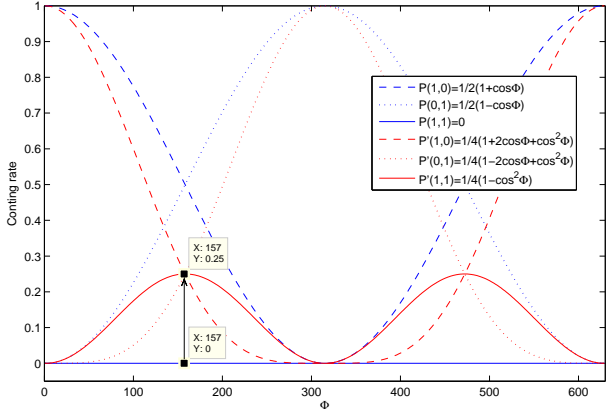}
\caption{Predictions of the local theory: $P(i,j)$, $i,j\in \{1,0\}$ label the probabilities when the detectors are timelike separated, and $P'(i,j)$ when they are spacelike separated (see text). For $\mathit \Phi=\pi/2$ the local theory predicts a change in the rate of coincident counts of: $P'(1,1)-P(1,1)=0.25$.}
\label{f3}
\end{figure}

The functions (\ref{P}) and (\ref{P'}) are plotted in Figure \ref{f3}.

Suppose now the two arms have the same length $L$ and the phase is set at value $\mathit\Phi=\frac{\pi}{2}$. The predictions of the local theory are:

- If detector D($-$) and D($+$) are timelike separated ($d\leq 0.75 m)$, coordinated firing behavior is possible, and the probabilities are given by (\ref{P}). Then  the counting rate at each detector is $P(1,0)=P(0,1)=0.5$, and the rate of coincident counts is $P(1,1)=0$.

-If D($-$) and (D$+$) are spacelike separated ($d>0.75 m$) coordinated firing behavior is thwarted and and the probabilities are given by (\ref{P'}). Then the counting rates are:  $P'(1,0)=P'(0,1)=0.25$ (1/2 of the times only one detector clicks), $P'(1,1)=0.25$ ($1/4$ of the times both detectors click jointly) $P'(0,0)=0.25$ ($1/4$ of the times no detector clicks).

In summary, the locality assumption implies that the rate of coincident counts  changes from $P(1,1)=0$ to $P'(1,1)=0.25$ when D($-$) is set beyond the signaling threshold (Figure \ref{f3}). This corresponds in (\ref{Pa}) to a phase shift of:
\begin{eqnarray}
\Delta\omega=\frac{\pi}{6}
\label{shift}
\end{eqnarray}

This shift (\ref{shift}) is the same as the represented in Figure \ref{f4} (see below).

By contrast, according to nonlocal quantum theory $P(1,1)=0$ is invariant under changes of the distance $d$, and no change in the interference pattern should be observed (negative result).

So, we have here a clear locality criterion allowing us to decide whether nature is nonlocal, provided the assumption of decision at detection.

\begin{figure}[t]
\includegraphics[width=80 mm]{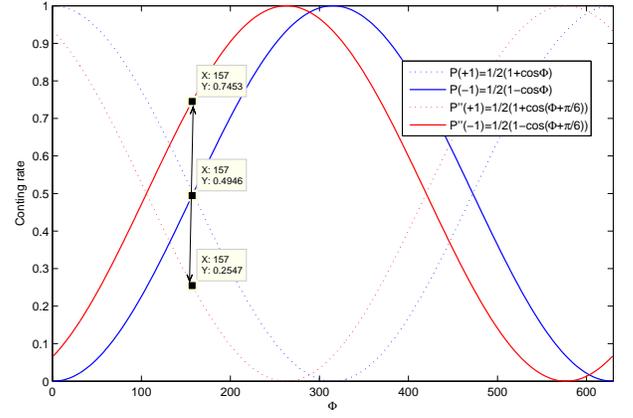}
\caption{Phase shift predicted by the ether theory in the proposed single-photon version of the Michelson-Morley experiment (see text). The shift corresponds to a change in the counting rate of detector D$(-)$ of: $P''(-1)-P(-1)=0.25$.}
\label{f4}
\end{figure}

Suppose now one uses the setup of Figure \ref{f2} to perform a single-photon version of the Michelson-Morley experiment.

According to the ether theory, in follow of the movement of the Earth around the Sun, the travel times of the light through the two interferometer arms exhibits a difference $\Delta t$ given by:

\begin{footnotesize}
\begin{eqnarray}
\Delta t = L \frac{v^2}{c^3}\approx\frac{L}{3} 10^{-16}
\label{MM1}
\end{eqnarray}
\end{footnotesize}

\noindent where one takes $v=30 km/s$ as the orbital velocity of the Earth \cite{mm}.

For photons of wavelength about $\lambda=900nm$ \cite{herald} the frequency $\nu$ is $\nu=\frac{c}{\lambda}=\frac{1}{3} \times\;10^{15}$. Taking account of (\ref{MM1}) one gets the phase shift:

\begin{footnotesize}
\begin{eqnarray}
\Delta\omega=2 \pi \Delta t \; \nu =2\pi\frac{L}{90}
\label{MM2}
\end{eqnarray}
\end{footnotesize}

To get the same  (\ref{shift}) as in the nonlocality experiment one has to set:
\begin{eqnarray}
L =7.5 m
\label{MM3}
\end{eqnarray}

Hence, if one uses an interferometer fulfilling (\ref{MM3}), when one of the arms is oriented in the direction of the motion of the Earth relative to the ether, the ether theory predicts that the counting rate of detector D($+$) changes from $P(+1)=0.50$ to $P''(+1)=0.25$, and that of D($-$) from $P(-1)=0.50$ to $P''(-1)=0.75$  (Figure \ref{f4}). This is equivalent to the change the local theory predicts for the rate of coincident counts $P'(1,1)-P(1,1)=0.25$ when detector D($-$) moves from timelike to spacelike separation.

Therefore the experiment is capable of ruling out locality with the same degree of reliability as it rules out the ether.

In the following we discuss some objections and highlight the interest of performing the proposed experiment.

\textbf{``One photon, one count''}. The local theory the experiment aims to test predicts that in $25\%$ of the detection events \emph{one single} photon produces two counts, and in other $25\%$ no count. This means that the energy is conserved in the average, but not in individual quantum processes. One may wonder whether it is worth to perform an experiment aiming to falsify such an assumption. I argue it is for two reasons:

1) As a matter of fact stochastic quantum dynamics may lead to the prediction that ``the energy is conserved on the average, but not in individual quantum processes'' \cite{ng}. In any case basic principles of physics should be based not only on intuition but also on experimental evidence. It is therefore worth to derive ``one photon one count'' from a possible negative result of the experiment with spacelike separated detectors sketched in Figure \ref{f2}, the same as one derives ``no-signaling'' from the negative result of the Michelson-Morley experiment.

2) If one takes beam-splitters as choice-devices (``pilot wave''), one can build local models that escape nonlocality at detection. These are the models tested in the conventional Bell experiments. However they are not less weird than the local theory assuming conservation of energy on the average. In fact such models are logically inconsistent, as I show in the following:

Local models assuming ``decision at the beam-splitters'' fulfill the well known locality criteria of Bell inequalities, and are refuted by the experimental violation of such inequalities \cite{jb}. However, such local models necessarily involve local hidden variables of the type of the ``empty pilot wave'' (otherwise they would not be able of accounting for interferences), which one cannot observe or detect directly: One can only characterize them by how the particles behave when observed. Ironically the local explanation bears the weird concept of entities existing and propagating locally in space-time that are unobservable and inaccessible in principle. But there is more: The motivation to oppose quantum nonlocality cannot be that of avoiding ``signaling'', since quantum nonlocality clearly does not induce such a thing. Actually, the motivation to oppose quantum nonlocality cannot be other than the wish to exclude from the physical reality entities that are not directly accessible to observation and control. This means that the local models tested by Bell experiments do assume in fact the type of entities they wish to refute. In this sense, they are logically inconsistent and, therefore, do not require experimental falsification. The relevance of the experimental violation of Bell inequalities is less that of demonstrating nonlocality than of implementing it to produce convenient cryptographic tools.

In summary, assuming ``decision at the beam-splitters'' bears absurdities, and hence the only nonlocality requiring experimental confirmation is nonlocality at detection.

\textbf{Loopholes}. Instead of drawing the conclusion of relativity, one could very well explain the Michelson-Morley negative result by stating that the source emits the photons with the hidden programm of traveling the arm in the direction of movement faster than the other one. How to refute such an interpretation? The only way to do it is by arguing with a gedanken-experiment: Suppose the experimenter sets the mirror at a distance $L/2$ and is able to move it to distance $L$ just before the photon arrives; then the source can no longer provide the particles with the correct programm.

Such an argument highlights that ``the freedom of the experimenter'' is a relevant issue also for relativity. One assumes that nature does not conspire: The source does not coax the experimenter to make settings according to the different ``hidden variables'' (velocities) the photons may carry. Conversely, one accepts as well that the free choices of the experimenter do not determine the velocity at emission in the past (no-retrocausation).

So, to accept the Michelson-Morley result one implicitly excludes that the source adapts the velocity of the emitted photons to the length of the path they have to travel, merely by invoking the ability of the experimenter to freely chose the settings, but without considering necessary to back the argument by a real experiment using ``settings chosen at random''. But in fact this is the same as endorsing the following standard for the scientific method.

\emph{Standard}: If some argument can be settled by a gedanken-experiment assuming the experimenter's freedom, it is not necessary to perform a real experiment where the experimenter chooses the settings at will.

Similarly, against the experimental demonstration of nonlocality at detection one could invoke sort of ``locality loophole'', already famous in the context of Bell experiments. In our case the objection reads: The detectors have time to become informed about the path lengths before the photons reach them, and can arrange a corresponding programm of firings. To meet this ``loophole'' one can again argue with a gedanken-experiment, by supposing that the experimenter changes at will the length $l$ so that the detectors don't have time to know it before the photons arrive.

However it is sheer impossible to implement this gedanken-experiment into a real one: On the one hand it is excluded that the experimenter himself chooses the settings at will fast enough. On the other hand, replacing the experimenter by some device generating random settings actually begs the question, since this means to accept freedom on the part of nature, and this is precisely a consequence of the feature one aims to demonstrate, i.e. (time-order independent) nonlocality \cite{as10}.

Thus, one meets here exactly the same problems as in the Michelson-Morley experiment, and applying the same \emph{Standard} one can conclude: If the proposed  experimental demonstration of nonlocality succeeds, it is as ``loophole free'' as that of relativity.

\textbf{The unity of locality and nonlocality}. The preceding analysis allow us to conclude that both relativity and quantum theory are based on the experimenter's freedom. This feature cannot be proved by scientific experiments since it precedes any coherent interpretation of the results. In this sense it should be considered an axiom of any theory.

On the basis of this axiom one accepts invariance of the light speed upon changes of the path length in the interferometer. Such invariance implies to give up the classical idea of ``particle'' as an entity well localized in space-time, since otherwise interferences would not be possible: Interferences require to introduce uncertainty in the time of emission and frequency bandwidth. As an interference experiment the Michelson-Morley relies on this quantum mechanical principles.

On turn, the negative result of this experiment implies the limit $v\leq c$ for ``signaling'', and highlights the necessity of doing the experiment proposed above (Figure \ref{f2}). The experiment makes clear that the standard view of decision at detection implies either nonlocality or an average conservation law of energy. In other words, nonlocality grants the principle that energy is conserved in each individual run.

Hence, nonlocality at detection appears as a reasonable and direct consequence of admitting interference phenomena and the photoelectric effect. Any theory that is based on the Michelson-Morley experiment, and in particular relativity, can't help accepting such nonlocality as well.

In this sense, Einstein's reluctance against nonlocality is somewhat astonishing: It may indicate that he was not quite aware of the role the photoelectric effect plays in the Michelson-Morley experiment, and therefore overlooked the oddities any local interpretation of interferences bears. All the more reason to really do the proposed single-photon version of the experiment.

\textbf{Conclusion.} An experiment has been proposed to demonstrate nonlocality at detection using a setup that can be used to reproduce the Michelson-Morley negative result as well. Both experiments happen under exactly the same conditions and both are supposed to falsify an equivalent prediction for changes in the counting detection rates. Thus, the demonstration of nonlocality can be considered as ``loophole free'' as that of relativity.

Additionally the experiment stresses that nonlocality at detection prevents locality from bearing odd concepts like ``conservation of energy only in average but not in each individual process'', and ``hidden local variables propagating in space-time but inaccessible to direct observation''. It also highlights that local models assuming ``decision at the beam-splitters'' are actually logically inconsistent.

Finally, the experiment clearly shows that relativity and quantum theory share the very same experimental basis, and derive from the same principles. Both seem to respond to the motivation of making a world characterized by the unity of local and nonlocal steering of detection outcomes. Relativity and quantum theory are two inseparable aspects of one and the same description of the physical reality. They do not have a ``frail peaceful coexistence'', but share a ``maximally entangled'' existence: we can't have one without the other.

This conclusion has been further strengthened in another paper \cite{as10a}.

Work to realize the proposed experiment is in progress.

\emph{Acknowledgments}: I am thankful to No\'{e} Curtz, Bruno Sanguinetti and Hugo Zbinden for ongoing work to prepare the setup, and Roger Colbeck, Nicolas Gisin and Renato Renner for discussions.

\end{document}